\documentclass [journal]{IEEEtran}
\usepackage{color,amsmath}
\usepackage {cite}

\usepackage{amssymb}
\usepackage{mathdots}
\usepackage[capitalise]{cleveref}
\usepackage{soul}
\usepackage[pdftex]{graphicx}
\graphicspath{{./figs/}{./picture/}}

\DeclareGraphicsExtensions{.pdf,.jpeg,.png}
\usepackage[caption=false,font=footnotesize]{subfig}

\begin{document}

\bstctlcite{IEEEexample:BSTcontrol}
\title{Wireless Localization with Space-Time Coded Reconfigurable Intelligent Surfaces}

\author{Mehdi Gholami, Soheil Khajavi, Mohammad Neshat, \IEEEmembership{Member, IEEE}, Simon Tewes, \IEEEmembership{Member, IEEE} and Aydin Sezgin, \IEEEmembership{Senior Member, IEEE}
\thanks{Manuscript received October 05, 2024. This work was supported in part by the German Federal Ministry of Education and
Research (BMBF) project 6G-ANNA [grant agreement number 16KISK095] and in part by the Deutsche Forschungsgemeinschaft (DFG, German Research Foundation) Project–ID287022738 TRR 196 (S03). \textit{(Corresponding author: Mohammad Neshat)}}
\thanks{Mehdi Gholami and Soheil Khajavi are with the  School of Electrical and Computer Engineering, College of Engineering, University of Tehran, Tehran, Iran}
\thanks{Mohammad Neshat is with the School of Electrical and Computer Engineering, College of Engineering, University of Tehran, Tehran, Iran, and also with the Centre for Wireless Innovation (CWI), Queen’s University Belfast, Belfast, UK (e-mail: m.neshat@qub.ac.uk)}
\thanks{Simon Tewes and Aydin Sezgin are with the Institute of Digital Communication Systems, Ruhr University Bochum, Bochum, Germany (e-mail: aydin.sezgin@rub.de)}}

\maketitle
\begin{abstract}
In this paper, a novel approach for wireless localization is proposed and experimentally validated that leverages space-time coded reconfigurable intelligent surfaces (RIS). It is demonstrated that applying proper single-bit codes to each RIS element, enables accurate determination of the direction of arrival (AOA) at the receiver. Moreover, we introduce different scenarios that such technique can be used for localization. By incorporating RIS, a passive component, the method significantly reduces the complexity found in previous localization techniques. Additionally, the use of 1-bit codes minimizes hardware requirements, offering a reliable, low-cost solution for localization in advanced telecommunications networks.
\end{abstract}


\section{Introduction}
Recent advancements in technology have markedly amplified the demand for innovation within the field of telecommunications. The forthcoming 6G technology seeks to enhance the telecommunications landscape by expanding service coverage, minimizing latency, and boosting the speed of data transmission \cite{tariq2020speculative}.

A critical aspect that necessitates enhancement, in conjunction with other advancements, is the precise localization of users within the network. An extensive investigation has been conducted to improve user localization in the upcoming generation \cite{de2021}.
A significant breakthrough in this investigation is the integration of intelligent components into the infrastructure.

The concept of reconfigurable intelligent surface (RIS) \cite{liu2021reconfigurable,elmossallamy2020reconfigurable,basar2019wireless,wu2019towards,alamzadeh2023,el2023,2aydin2023IEEE,alamzadeh20,aydin2023IEEE} is considered to be a solution and an important evolution in addressing these requirements. RIS is noted as a reconfigurable metasurface consisting of a 2D-array of elements. By controlling amplitude and phase of the reflected (transmitted) wave from each element, it would be possible to control the wavefront. RIS can play a significant role not only in improving telecommunication networks but also as a tool for imaging and sensing. Recently, there has been studies on the application of RIS in localization \cite{wymeersch2020radio,wymeersch2020beyond,ma2020indoor,zhang2021metalocalization,elzanaty2021reconfigurable,liu2021,zhang2022toward,aydinlocalization2021}. It has a significant improvement in localizing the users who are not within direct line of sight, e.g. obstructed by objects in the environment. Recently, Cao et al. \cite{cao2023} presented the idea of using a 1-bit code on a time modulated refectory metasurface for estimation of the direction of arrival.
This paper expands this novel approach to experimental measurements in the field of wireless localization  using space-time coding RIS. Our methodology is built upon our previous idea on the vector representation of the binary codes
\cite{gholami2022a}. We employ single-bit binary codes as a sequence of $L$ binary bits, consisting of 0s and 1s to allocate one of two states (ON/OFF) to the RIS elements within a column of a 2D-RIS, with a one-bit shift applied to the adjacent RIS-columns. Through the utilization of this technique, several harmonics are generated, each propagating on a discernible direction. 
Each harmonic possesses a unique spatial and temporal frequency, which can be measured to ascertain the direction of the reflected harmonic. 
Ultimately, we experimentally assess our proposed approach in both an anechoic chamber and a practical real-world setting.

The subsequent sections of the article are structured as follows. 
In section \ref{Fundumentals}, the principles underlying localization through space-time coded RIS is presented. The concept of Fourier vector representation of binary codes and the generation of distinct harmonics along with the theoretical basis for estimating angle-of-arrival (AoA) are discussed.
In section \ref{method}, the methods to assess AoA from the harmonics and various scenarios for localization is detailed. 
In section \ref{measurements}, we present the experimental measurements conducted to validate our proposed method a real-world environment, followed by concluding remarks in section \ref{conclusion}.

\section{Fundamentals of Localization with Space-Time Coding} \label{Fundumentals}
In our previous work \cite{gholami2022a}, we demonstrated that by applying sequencing iterative codes to the elements of a metasurface, various harmonics can be generated. We also demonstrated that different codes control the frequency, phase, and amplitude of the wave reflected from each element. This indicates that the total generated harmonics by the RIS are dependent on the binary codes applied to each element.

\begin{figure}[t]
\centering
\subfloat[]{\includegraphics[height=1.5in,page=1]{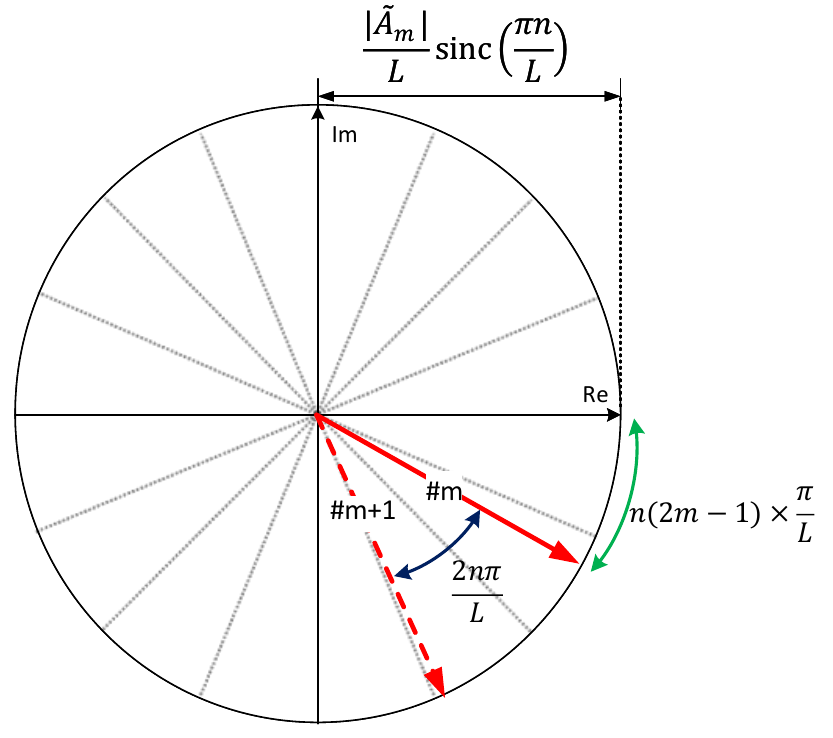} \label{1.1}}
\subfloat[]{\includegraphics[height=1.45in,page=1]{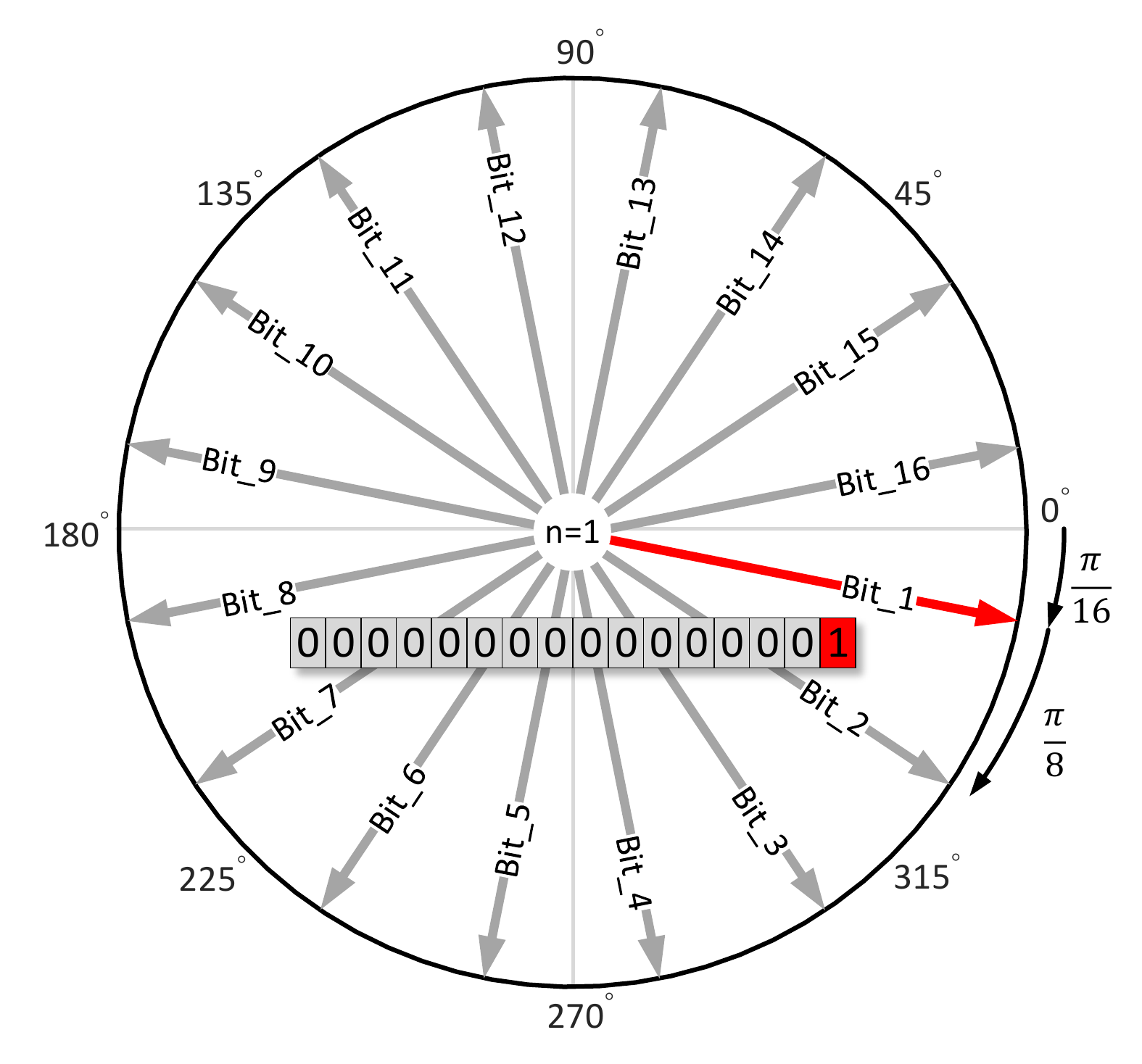} \label{1.2}}
\caption{ 
\protect\subref{1.1} General form of Fourier vector representation for $n^{th}$ harmonic due to a code with only $m^{th}$ bit being set to 1. By one-bit shift, the $n^{th}$ harmonic experiences a phase shift of $2n\pi/L$. \protect\subref{1.2} Representation of the first harmonic due to a 16-bit code with its first bit being set to 1. The first harmonic undergoes a phase shift of $\pi/8$ by each one-bit shift.}
\label{vector_representation}
\end{figure}

Our proposed methodology for localization through space-time coding hinges on specific codes that encompasses components in all n harmonics. Utilizing vector representation \cite{gholami2022a} enables the selection of such codes with components that extend across all frequency harmonics. To illustrate phase manipulation through switching with vector representation, a time-harmonic function, $\exp(j2\pi f_c t)$, is assumed and multiplied by a switching function, $\tilde{S}(t)$. In the frequency domain, the phase distribution of the waveform is the same as the switching function, $\tilde{S}(t)$, shifted by the carrier frequency, $f_c$. The switching function can be considered as a time sequence of repeating digital bits which are known as a binary code. The switching function in the time domain can be represented as:
\begin{equation}
\label{eq.swtch_fun}
\tilde{S}\left(t\right)=\sum_{m=1}^L\tilde{A}_mU_m\left(t\right);~~ 0<t<T_0 ~,
\end{equation}
where $\tilde{A}_m$ is typically a complex value that modulates the amplitude and phase of the time-harmonic wavefront in each bit duration. The term $U_m(t)$ represents a periodic pulse function corresponding to the $m^{\textrm{th}}$ bit of the code and is defined within each period, $T_0$, as:
\begin{equation}\label{eq.U_m}
  U_m\left(t\right)=\left\{\begin{array}{l}1,\ \ \ \ \left(m-1\right)\tau{}\leq{}\
t\leq{}m\tau{} \\
0,\ \ \ \ \ \ \ \ \ \ \ \ \ \ \ \ \ \ \textrm{o.w.}\ \end{array}\right.,
\end{equation}
\noindent where $\tau = T_0/L$ represents the time duration of each bit.

By Expanding this function into Fourier series, the amplitude and phase of the $n^{\textrm{th}}$) harmonic of $f_0=1/T_0$ can be determined by \cite{RN57}
\begin{subequations}
\label{eq.S}
\begin{align}
  \label{eq.S.amp}
  S_n&=\left\vert{}\sum_{m=1}^L\frac{\tilde{A}_m}{L}\textrm{sinc}\left(\frac{\pi{}n}{L}\right)\exp{\left[-\frac{j\pi{}n\left(2m-1\right)}{L}\right]}\right\vert{},
 \\
  \label{eq.S.pha}
\Theta_n&=arg\
\left\{\sum_{m=1}^L\frac{\tilde{A}_m}{L}\textrm{sinc}\left(\frac{\pi{}n}{L}\right)\exp{\left[-\frac{j\pi{}n\left(2m-1\right)}{L}\right]}\right\},
\end{align}
\end{subequations}
\noindent where $arg\{.\}$ denotes the phase. Upon further investigation of (\ref{eq.S}), it becomes apparent that each term in the summation may be regarded as a complex phasor that can be represented by a vector in the complex plane. Each specific bit of the code can be associated with a single vector. Therefore, a complete code can be presented as a sum over $L$ vectors. In this case, $\tilde{A}_m$ is treated as binary ($\tilde{A}_m \in {0,1})$. The general form of Fourier vector representation of two codes with the $m^{th}$ and ${m+1}^{th}$ bit being set to 1 is illustrated in Fig. \ref{1.1}.

Every single bit shift induces a phase difference of $\Delta \phi_n$ for $n^{th}$ harmonic as:
\begin{equation}
\Delta \phi_n=\frac{2 n \pi}{L}.
\label{eq_2}
\end{equation}

The Fourier vector representation of a single-bit code of length 16, with each bit independently set to 1, is shown in Figure \ref{1.2}. This figure demonstrates that only a single-bit code allows for shifting without any overlapping in the harmonics. If a code with multiple bits set to 1 is used, shifting may cause the vectors to overlap, leading to interference between different harmonics. A single-bit code, however, eliminates the risk of destructive interference between harmonics and ensures the distinct generation of each harmonic.

By applying single-bit codes to the RIS elements and shifting the codes by one bit column-wise, it resembles a phased array with uniform spacing of $d$ and a phase difference as given in \eqref{eq_2} between neighboring columns for $n^{th}$ harmonic. Therefore, the steering angle of the main lobe emanating from the RIS for the $n^{th}$ harmonic, $\theta_n$, is calculated as:

\begin{equation}
\theta_n = \operatorname{asin}\left(\frac{n\lambda}{Ld}\right).
\label{eq_3}
\end{equation}

\begin{table}[b]
\caption{Calculated steering angles for 8 harmonics with $L=16$ and $d=\lambda/2$.}
\centering
\begin{tabular}{|c|c|c|c|c|c|c|c|}
\hline
$n$ & $\theta_n$  & $n$ & $\theta_n$  & $n$ & $\theta_n$  & $n$ & $\theta_n$  \\
\hline
1 & $7.18^\circ$ & 2 & $14.48^\circ$ & 3 & $22.02^\circ$ & 4 & $30.00^\circ$  \\
\hline
5 & $38.68^\circ$ & 6 & $48.59^\circ$ & 7 & $61.04^\circ$ & 8 & $90.00^\circ$  \\
\hline
\end{tabular}
\label{steering_angle}
\end{table}

From \eqref{eq_3}, it is clear that each harmonic emanates from the RIS with a specific steering angle. Negative order harmonics are steered symmetrically along the negative angles. The zero-order harmonic also reflects at the zero angle that is normal to the RIS surface.

In order to maintain symmetry, we choose the number of bits to be equal to the number of columns in the RIS. In the analysis, we examine a 16-bit code for a $16\times16$ RIS, where the stearing angles for eight harmonics is given in Table \ref{steering_angle}. 

\begin{figure}[t]
\centering
\subfloat[]{\includegraphics[width=3in,page=1]{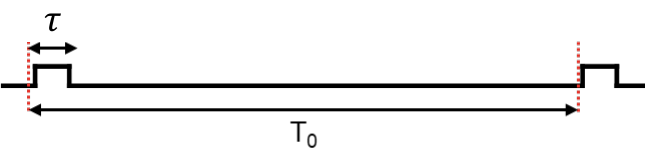} \label{2.1}} \\
\subfloat[]{\includegraphics[width=3in,page=1]{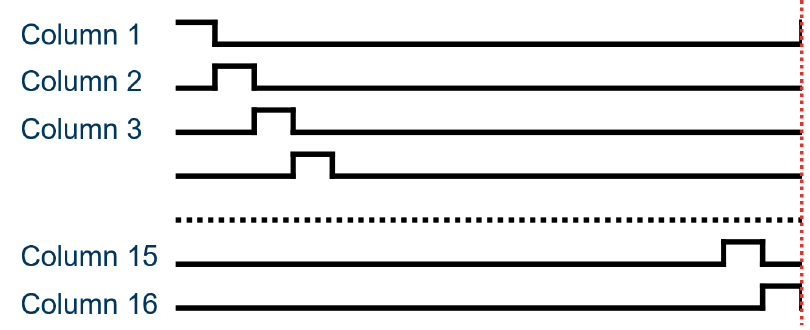} \label{2.2}}
\caption{ 
\protect\subref{2.1} Signal code with only one bit at ON state.
\protect\subref{2.2} Signal codes applied to the elements of the RIS column-wise where each code is shifted by one bit as compared to the next.}
\label{Bias}
\end{figure}

Fig. \ref{Bias} demonstrates the signal codes applied to the columns of the RIS. The duration of one bit and one code is $\tau_0$ and $T_0=16\tau_0$, respectively. Harmonics in the frequency spectrum are separated apart by the value of $f_0=1/T_0$. The hardware used in this research has a switching speed of $\tau_0 = 1.87$  ms and $T_0 = 30$ ms, which leads to a first harmonic frequency of $f_0 = 33.427$ Hz.

\begin{figure*}[ht]
\centering
\includegraphics[width=7in]{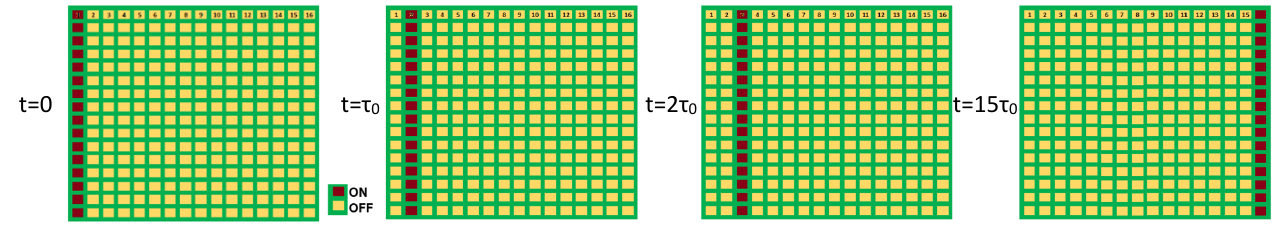}
\caption{The ON-OFF state of each element in the RIS in different time frames.
} 
\label{3}
\end{figure*}

Fig. \ref{3} shows the binary state of the elements in the RIS. With each time step, all the elements in a single column are set to state 1 (ON), while the rest of the elements are in state 0 (OFF), and in the subsequent time step, the column sets back to 0 while the adjacent column becomes 1. This proceeds until the last column, and it consistently repeats as long as the localization is in progress. By uniformly changing the states of the elements in every column, the harmonics are steered in the horizontal direction. Implementing such space-time coding of RIS row-wise  would lead to beam steering in elevation direction.

\begin{figure}[ht]
\centering
\includegraphics[width=3.5in]{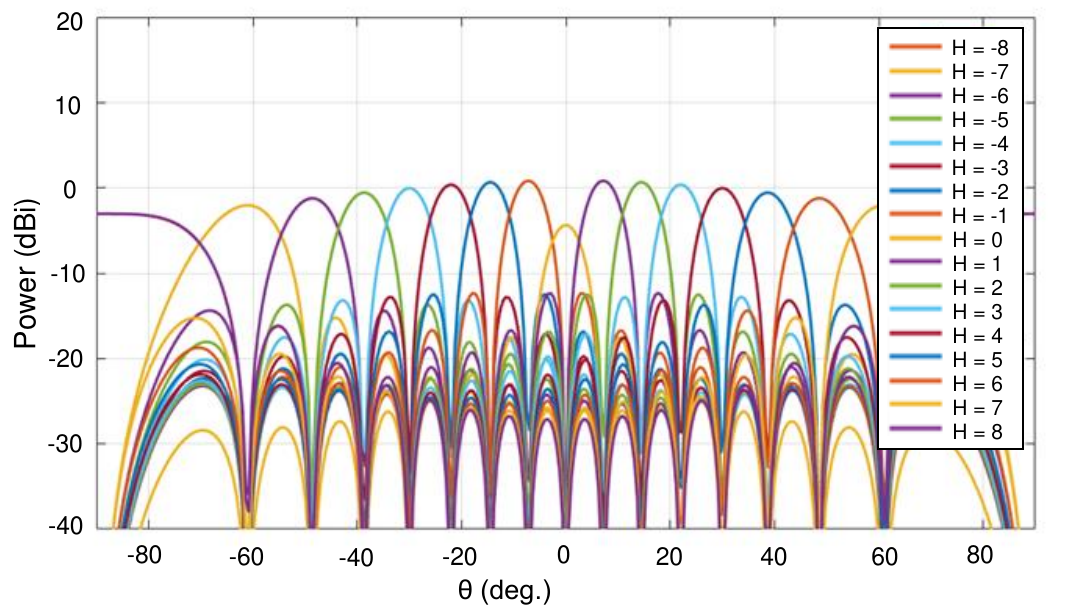}
\caption{Simulated radiation pattern of reflected harmonics from the RIS when a pilot beam is incident normal to the RIS, and the elements are coded uniformly column-wise as shown in Fig. \ref{3}. H is the number of the reflected harmonics.
} 
\label{4}
\end{figure}

Fig. \ref{4} shows the simulated radiation pattern of reflected harmonics ($-8\leq n\leq 8$) from the RIS when a pilot beam is incident normal to the RIS, and the elements are coded uniformly column-wise as shown in Fig. \ref{3}. As expected, each harmonic has a main lobe in a specific direction consistant with those calculated in Table \ref{steering_angle}, and all of these harmonics cover $180^{\circ}$ field of view in front of the RIS. In this simulation, we employed (0, $\pi$) binary phase states of reflection coefficient for elements, that resulted in a reduced amplitude of the zero-order harmonic ($n=0$) emanating normal from the RIS. 

\begin{figure}[t]
\centering
\includegraphics[width=3.5in]{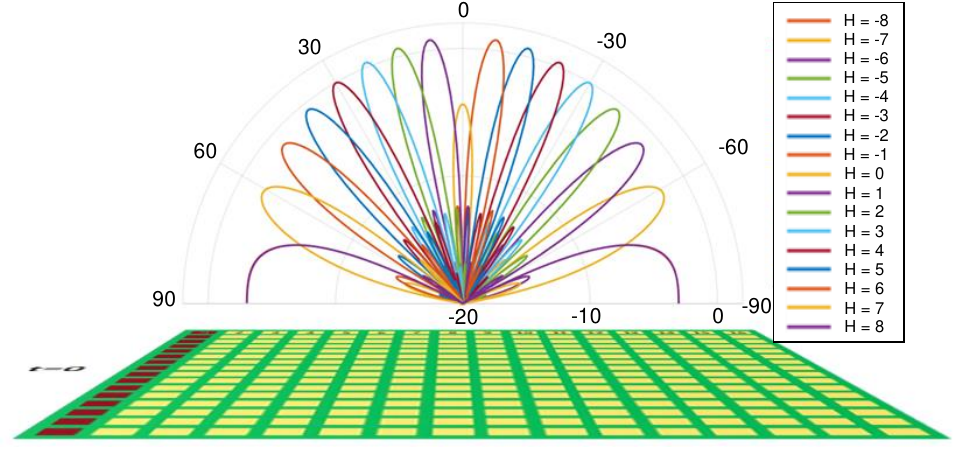}
\caption{Displaying the radio coverage of the harmonics in the space facing the RIS.
} 
\label{5}
\end{figure}
The radio coverage of the harmonics in the front side of the RIS for our suggested technique is shown in Fig. \ref{5}. As anticipated, every harmonic exhibits peak power at a certain angle. The specification of maximum power in the harmonics can serve as a reference point for determining the angle-of-arrival. The technique commences with the sampling of the received wave and the determination of the amplitude of the harmonics. This enables an accurate estimation of the AoA with respect to the RIS.

\begin{figure}[ht]
\centering
\includegraphics[width=3.5in]{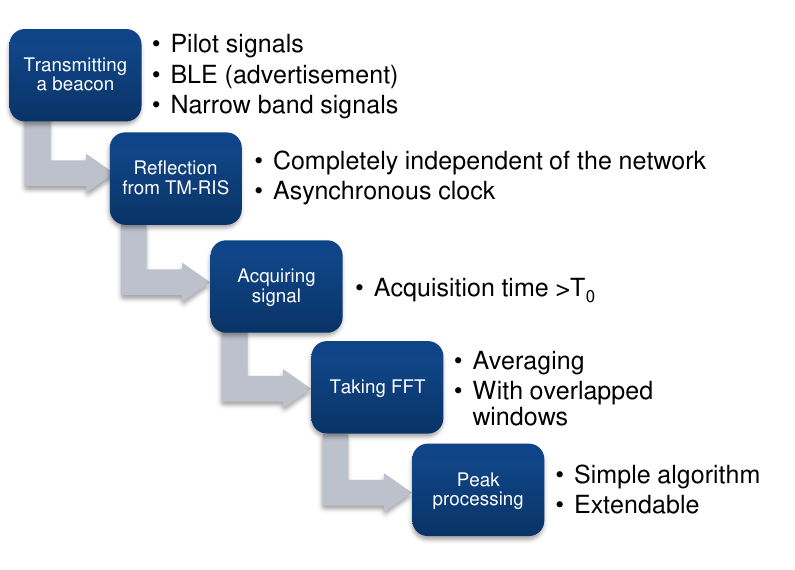}
\caption{The sequential steps of applying the proposed method for localization using space time coding RIS. 
The first step involves beacon transmission, which can be a pilot signal, Bluetooth Low Energy (BLE), or any generated narrow-band signal. In the second step, the signal is reflected by the time-modulated RIS, which is dynamically activated by codes through an asynchronous clock. Next, a moving average is applied to overlapping windows after performing a Fourier transform. Finally, an algorithm identifies the peak corresponding to the Angle of Arrival (AoA).} 
\label{6}
\end{figure}

\begin{table*}[th]
\caption{Different scenarios for localization in an advanced wireless network using the proposed method.}

\centering

\begin{tabular}{|p{1.25in}|p{2.7in}|p{1in}|}
\hline
\multicolumn{1}{|c|}{\textbf{Assumptions}} & \multicolumn{1}{|c|}{\textbf{Scenarios}} & \multicolumn{1}{|c|}{\textbf{Configuration}} \\
\hline
\textbf{Known:} \hfill \break Location of the base station \hfill \break Location of the RIS \hfill \break
\textbf{Unknown:} \hfill \break Direction of the user & 
The base station generates the wave with the main frequency. The user provides feedback on the magnitude of the received harmonics to the network. The network uses the position of the RIS, base station and the power of received harmonics to predict the user's direction.
& \raisebox{-\totalheight}{\includegraphics[width=1in]{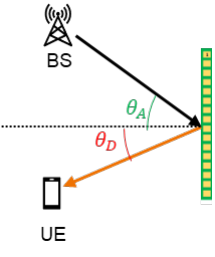}} \\

\hline
\textbf{Known:} \hfill \break Location of the RIS \hfill \break
\textbf{Unknown:} \hfill \break Direction of the user & 
The user equipment seeks to determine the angle between itself and the RIS. In order to accomplish this, it produces a single tone wave and evaluates the magnitude of the reflected harmonics, thereby predicting the its angle to the RIS. 

& \raisebox{-\totalheight}{\includegraphics[width=1in]{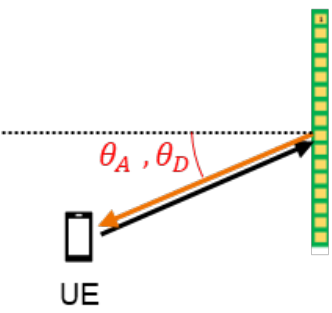}} \\
\hline
\textbf{Known:} \hfill \break Location of the base station \hfill \break
\textbf{Unknown:} \hfill \break Direction of the RIS & 
The base station aims at determining the direction of the RIS with respect to itself. It emits a single tone wave and evaluates the reflected power of the harmonics. By taking into account the power of the harmonics, it is able to estimate the angle. 

& \raisebox{-\totalheight}{\includegraphics[width=1in]{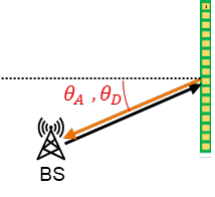}} \\
\hline
\textbf{Known:} \hfill \break Location of the base station \hfill \break Location of the RIS \hfill \break
\textbf{Unknown:} \hfill \break Location of the user & 

By employing multiple RISs, it is feasible to ascertain the precise coordinates of the user. Each RIS can operate at a distinct frequency. We can utilize the gap between harmonic frequencies to save the spectrum, enabling simultaneous operations and efficient localization. 

& \raisebox{-\totalheight}{\includegraphics[width=1in]{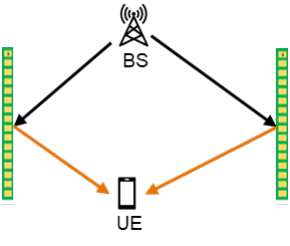}} \\
\hline
\textbf{Known:} \hfill \break Location of the base station \hfill \break Location of the RIS \hfill \break
\textbf{Unknown:} \hfill \break Location of the user & 

By combining the time difference of arrival (TDOA) \cite{keykhosravi2021} with localization using RIS, it becomes feasible to accurately ascertain the precise location of the user.

& \raisebox{-\totalheight}{\includegraphics[width=01in]{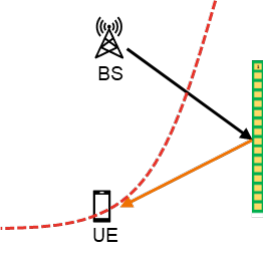}} \\
\hline
\end{tabular}

\label{table_scenarios}
\end{table*}

\section{Proposed methods to Attain Localization through Space-Time Coded RIS} \label{method}

In the preceding section, we demonstrated that through utilization of a single tone pilot that illuminates the RIS, and analyzing the reflection pattern with distinct steering angle for each harmonic it is possible to predict the relative directions of the components. We propose to use such characteristic for localization in the upcoming era of wireless communication through the steps illustrated in Fig. \ref{6}.

Within telecommunications networks, a specific duration of time is designated for the purpose of determining the direction and position of users. In this time frame, network elements participate in the exchange of information that results in direction-finding and location-tracking of a user \cite{laoudias2018}. In the proposed approach, the first stage involves the transmission of a single tone pilot, a group of pilots, or even a narrow-band signal sent omnidirectionally. The RIS in this system can operate independently from the network and does not require frequency or time synchronization with other network elements. Every user within the coverage of the RIS, samples a short time interval of the main signal and its harmonics with a sampling duration exceeding $T_0$. The frequency spectrum of the received signal is determined after sampling. Since the frequency spectrum may change and distort due to environmental factors and channel variations, the average over the Fourier transform of moving time frames of the signal is calculated.

After calculating the average over the frequency spectrum, it is necessary to measure the magnitude of each harmonic. If there is no information about the harmonic frequencies at the receiver, correlation methods or peak searching can be used to precisely locate and determine their values. The magnitudes of the harmonics provide essential information for estimating the AoA relative to the RIS through a series of computations. First, the magnitude of each received harmonic is multiplied by its corresponding harmonic in the radiation pattern shown in Fig. \ref{4}. Subsequently, the radiation patterns are summed, resulting in a single peak within the combined pattern, which yields the estimated angle.
With deriving AoA, various scenarios can be suggested for direction-finding and exact location-tracking in RIS-assisted telecommunications networks. Some of these scenarios are outlined in Table. \ref{table_scenarios}. It is important to note that  determining the user's exact location requires the use of multiple RISs that operate at distinct frequencies, enabling simultaneous operations.\\

\section{Measurement Results on Wireless Localization using Space-Time Coding RIS} \label{measurements}
\begin{figure}[ht]
\centering
\includegraphics[width=3.5in]{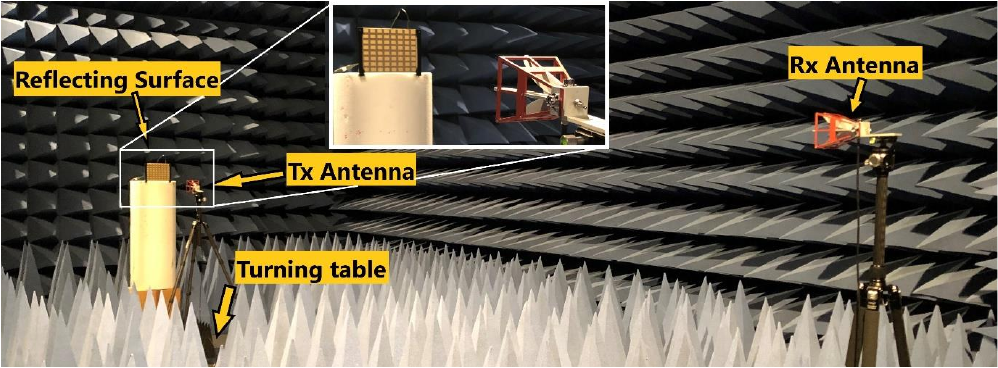}
\caption{Characterization setup of the space-time coding RIS in an anechoic chamber.
} 
\label{antenna_room}
\end{figure}
\begin{figure}[t]
\centering
\subfloat[]{\includegraphics[width=3.3in]{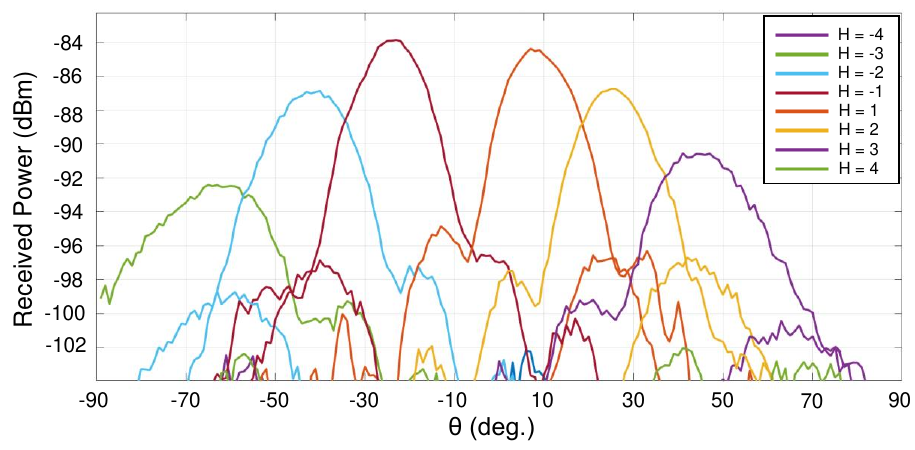} \label{9}} \\
\subfloat[]{\includegraphics[width=3.3in]{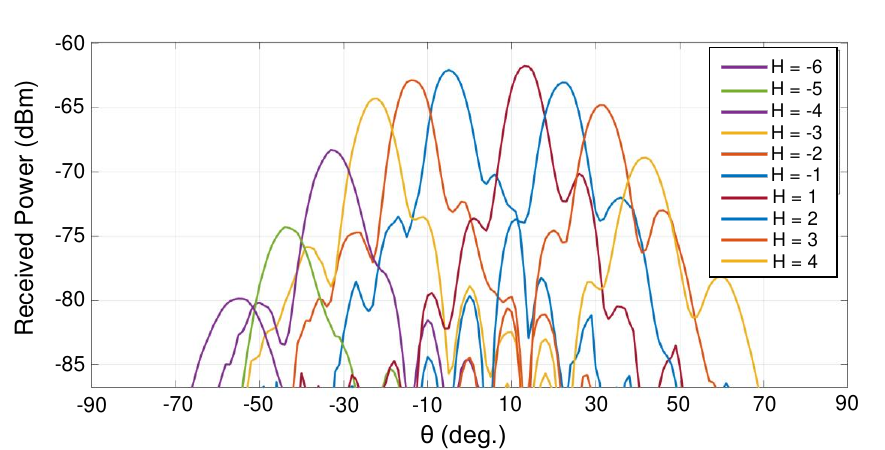} \label{10}}
\caption{ 
Measured radiation pattern of the harmonics with a \protect\subref{9} $8\times8$ and 
\protect\subref{10} $16\times16$ elements RIS. H is the number of harmonics. The coding sequence of RIS is illustrated in Fig. \ref{3}}.
\label{9_10}
\end{figure}

This section focuses on assessing the accuracy of the proposed localization approach by conducting a series of practical experiments. We examined two in-house fabricated RISs, having arrays of $8\times8$ and $16\times16$ elements. The applied code to $8\times8$ and $16\times16$ RIS has 8 and 16 bits, respectively. We first measured the radiation pattern of the harmonics through a characterization setup in an anechoic chamber as shown in Fig. \ref{antenna_room}. For this measurement, the RIS along with a TX antenna are installed on a same rotating table. The TX antenna sends a single tone pilot wave toward the space-time coding RIS, and the reflected harmonics are detected by a fixed RX antenna as shown in Fig. \ref{antenna_room}. The rotary table makes a short pause at each angular position, during which the received signal is sampled, and the table rotates to the next angle. By processing all received signals from each angle, their frequency spectrum are extracted, and the magnitude of each harmonic with respect to angular positions are recorded. Subsequently, the radiation patterns of all harmonics are plotted  as shown in Fig. \ref{9_10}.

As it is evident from Fig. \ref{9}, harmonic orders of $n= -3$ to $n=+3$ are dominant in the radiation, allowing one to disregard the remaining harmonics. These six harmonics cover a $180^{\circ}$ field-of-view in front of the RIS. The harmonic patterns are tailored such that at each angle only one dominant harmonic exists (except for intersecting points). Additionally, for a more accurate estimation of the Angle of Arrival (AoA), it is beneficial to consider the relative magnitudes of neighboring harmonics, instead of focusing exclusively on the dominant harmonic.

As anticipated, enlarging the size of the RIS results in a narrower harmonic beamwidth (see Fig. \ref{10}), hence improving the angular resolution. It is expected to obtain higher angular resolution as the size of the RIS increases.

It is observed from Fig. \ref{9_10} that the magnitudes of the harmonic reduce as the reflecting angle increases. This is due to the reduction in the effective area of the RIS at larger angles.

A valid question can be that how many harmonics would dominantly radiate away  from the RIS? 

Given that the effective cross-sectional area, the proportion of wave that is intercepted by the RIS, is nearly zero for this harmonic at $\theta_{n_{max}}=90^\circ$ and according to \eqref{eq_3},
\begin{equation}
\frac{\pi}{2} = \operatorname{asin}\left(\frac{n_{max}\lambda}{Ld}\right) \xrightarrow{d=\lambda/2} n_{max}=\frac{L}{2},
\label{eq_6}
\end{equation}
the largest dominant harmonic order that leads to a real-value steering angle is $n_{max}=L/2$ (where $L$ is the length of the binary code or, in other words, the number of columns in our utilized RIS and $d=\lambda/2$). Further harmonics can be ignored due to their non-radiating behaviour. Consequently, for a RIS with $L$ columns, in an optimal scenario, only $L$ unique beams are present. Depending on the type of AoA estimation algorithm (which may be a subject of future research), one can divide the magnitude level of each beam uniquely into three partitions by considering neighboring harmonics: only one strong harmonic, a combination of one strong harmonic and an adjacent harmonic on the left, and a combination of one strong harmonic and an adjacent harmonic on the right. Therefore, approximately $3L$ angular partitions can be made in front of the RIS. This means that if a RIS consists of 60 columns, there would be $\sim1^\circ$ angular resolution for AoA estimation.

\begin{figure}[b!]
\centering
\subfloat[]
{\includegraphics[width=2.8in,page=1]{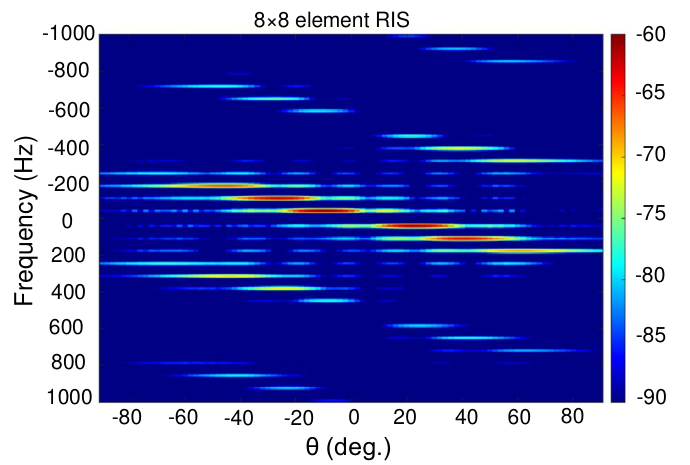} \label{11.2}}\\
\subfloat[]
{\includegraphics[width=2.8in,page=1]{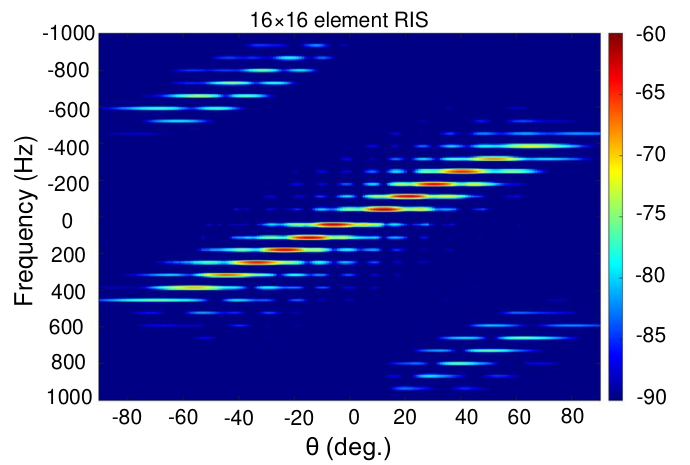} \label{11.1}} 
\caption{Measured colormap of radiating harmonic power in frequency-angle plane for a space-time coding RIS with \protect\subref{2.1} $8\times8$ and \protect\subref{2.2} $16\times16$ elements. Colorbar is in dBm unit.}
\label{11}
\end{figure}

Fig. \ref{11} shows the measured colormap of radiating harmonic power in frequency-angle plane for a space-time coding RIS with $8\times8$ and $16\times16$ elements. As seen in Fig. \ref{11}, only the predominant harmonics within the range of $n=-L/2$ to $n=+L/2$ are present in front of the RIS, although the highest harmonic power is quite weak.

\begin{figure}[t]
\centering
\includegraphics[width=3.1in]{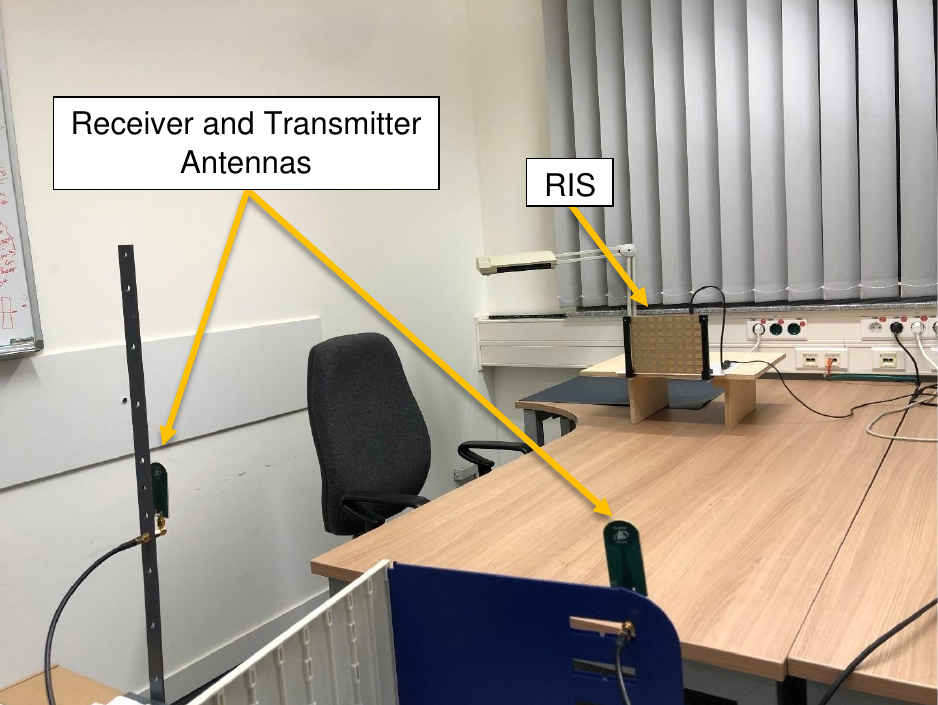}
\caption{
Testing the proposed localization method using the space-time coding RIS in an office space.
} 
\label{12}
\end{figure}

Measurements in the anechoic chamber demonstrate completely separate radiation patterns for each harmonic, but in a real-world setting, due to high interferences and secondary reflections, the measured radiation patterns do not match the quality obtained in the anechoic chamber. Fig. \ref{12} illustrates an experimental arrangement of components in a typical indoor workspace. Another part of the room not shown in the image includes a computer, bookshelf, and whiteboard. 

In these measurements, the transmitter continuously emits a single tone pilot in normal direction to the RIS at 5.385 GHz. The RIS elements are coded similar to the previous experiments. The receiver is positioned at various angles relative to the normal direction to the RIS.

\begin{figure}[h]
\centering
\includegraphics[width=3.3in]{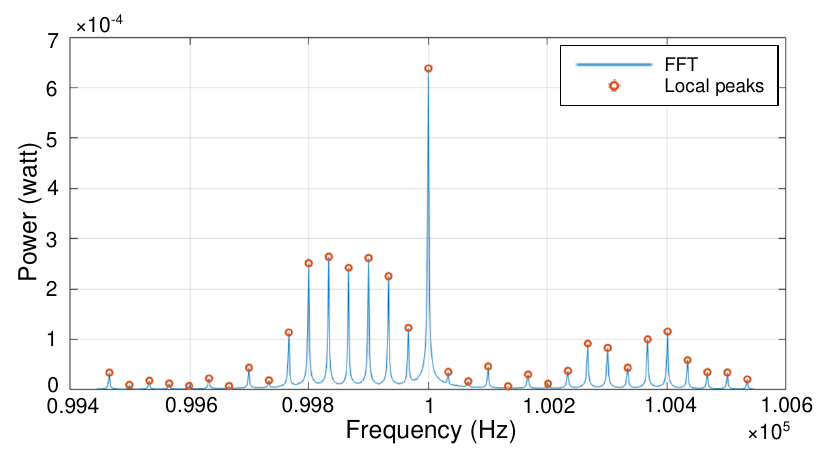}
\caption{Measured frequency spectrum of the received signal at a specific angle in the test setup shown in Fig. \ref{12}. Detected harmonic frequencies are marked with red circles.
} 
\label{13}
\end{figure}

\begin{figure}[h!]
\centering
\includegraphics[width=3.3in]{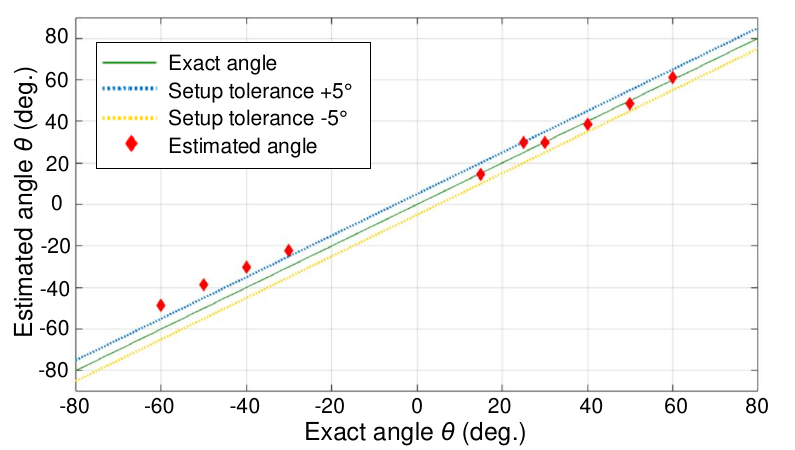}
\caption{The result of estimating the receiver's angle using the proposed space-time coding RIS.
} 
\label{14}
\end{figure}

The frequency spectrum of the received signal by the receiver at one of the test points is shown in Fig. \ref{13}. The application of averaging in the computation of the frequency spectrum creates favorable conditions for identifying the harmonics with high distortion. By comparing the relative harmonic magnitude with the characterized harmonic power map (see Fig. \ref{11}), an estimation of the receiver's angle relative to the RIS can be obtained.

Fig. \ref{14} illustrates the results of estimating the receiver's angle using the proposed space-time coded RIS. As previously mentioned, the power of the harmonic peaks is multiplied by their corresponding harmonics in the radiation pattern and then summed. A dominant peak at a specific angle is used to indicate the AoA.
The red marks represent the estimated angles where the receiver was positioned. The green line indicates the range where the actual angle matches the estimated angle perfectly. The yellow and blue lines also indicate $\pm5^{\circ}$ error. As seen, the estimation results closely match the actual values. Only at negative angles there is an offset, which could be due to the uncertainty in the placement of the experimental components. It can be observed that even at these negative angles, the red points follow a line, indicating a systematic error in the experiments. For such an arrangement with $N=8$ columns, we predicted 24 discernible angular partitions in front of the RIS, indicating an angular resolution of $7.5^\circ$. This prediction aligns well with the measurement results, although further optimization could lead to better accuracy.

\section{Conclusion}
\label{conclusion}
Our study underscores the promising role of space-time coding arrays with RIS in enabling precise and efficient localization capabilities for future wireless communication systems. This technology opens avenues for innovative applications in direction-finding, location-tracking, and enhancing overall network performance. Future research endeavors could focus on refining AoA estimation algorithm, expanding experimental validations, and exploring practical implementations to realize the full potential of this emerging technology in telecommunications.

\bibliographystyle {IEEEtran}
\bibliography{refs}

\end{document}